\documentclass[numreferences]{kluwer}

\usepackage{amsmath}
\usepackage{multirow}
\def\half{{\textstyle {1\over 2}}}
\def\thrd{{\textstyle {1\over 3}}}

\usepackage{graphicx}

\begin{document}

\begin{opening}

\title{Surfactant Spreading on a Thin Liquid Film: Reconciling Models and
Experiments}

\author{Ellen R. \surname{Swanson}}
\institute{Dept. of Mathematics,  Centre College, Danville, KY 40422}
\author{Stephen L. \surname{Strickland}}
\institute{Dept. of Physics,  NC State University, Raleigh, NC 27695 }
\author{Michael \surname{Shearer}}
\institute{Dept. of Mathematics, NC State University, Raleigh, NC 27695}
\author{Karen E. \surname{Daniels}}
\institute{Dept. of Physics,  NC State University, Raleigh, NC 27695 }

\begin{abstract}
The spreading dynamics of surfactant molecules on a thin fluid layer is of both
fundamental and practical interest. A mathematical model formulated by
Gaver and Grotberg \cite{Gaver-1990-DLS} describing the spreading of a single layer of
insoluble surfactant has become widely accepted, and several experiments on
axisymmetric spreading have confirmed its predictions for both the  height profile of the free
surface  
and the spreading exponent (the radius of the circular area covered by
surfactant grows as $t^{1/4}$). 
However, these prior experiments have primarily utilized surfactant quantities
exceeding (sometimes far exceeding) a monolayer. In this paper, we report that
this regime is characterized by a mismatch between the 
timescales of the experiment and model, and additionally find
that the spatial distribution of surfactant molecules differs substantially
from the model prediction. For experiments performed in the monolayer regime
for which the model was developed,
the surfactant layer is observed to have a spreading exponent of less than
$1/10$, far below the predicted value, and the surfactant distribution is
also in disagreement. These findings suggest that
the model is inadequate for describing the spreading of
insoluble surfactants on thin fluid layers. 
\end{abstract}

\end{opening}

\section{Introduction \label{s:intro}}

Axisymmetric spreading of an insoluble surfactant on a thin layer of
incompressible fluid has been the subject of numerous experimental and
mathematical studies \cite{Bull-1999-SSS, Dussaud-2005-SCI, Fallest-2010-FVS, Gaver-1990-DLS, Jensen-1992-ISS, Jensen1994}.
 Motivated by the biomedical application of aerosol medications delivered to the
thin
fluid lining the lung, Gaver and Grotberg \inlinecite{Gaver-1990-DLS}
derived a mathematical model, based on lubrication theory, that
couples the height profile of the fluid surface $h(r,t)$ to the local surfactant
concentration $\Gamma(r,t)$. This model captures the driving force
associated with the Marangoni surface stress induced by spatial variations in
surfactant  concentration, which in turn depends on an equation of state that
specifies the dependence of surface tension $\sigma$ on  $\Gamma$.
While the model was developed
for monolayer applications of surfactant, it has come to be applied both
above \cite{Dussaud-2005-SCI} and near \cite{Bull-1999-SSS}
the critical monolayer concentration $\Gamma_c$, the concentration above which
a single layer of surfactant molecules can no longer form.
Similar models have been used to study 
thin films in bronchial systems \cite{Grotberg2011},
ocular systems including blinking dynamics \cite{Braun-2012-DTF}, bulk solute
transport \cite{Engineering1994},
drying of latex paint \cite{Evans2000,Gundabala2006,Gundabala2008},
ink-jet printing \cite{Hanyak2011}, and
secondary oil recovery \cite{Sinz2011b,Hanyak2012}.

Numerical simulations have been used to confirm several predictions of the model
\cite{Jensen-1992-ISS,Peterson-2010-RSS} based on analysis of self-similar
solutions. The simulations  have also yielded detailed 
information about the spatiotemporal evolution of the free surface height and
surfactant concentration profiles. One key observation is that the motion of the
leading edge of the surfactant-covered region sets its spreading rate. The
decrease in 
concentration at the leading edge induces a surface stress (Marangoni stress),
which drives a capillary ridge  (local maximum) in  the fluid free surface that
propagates along with the leading edge. In spite of these advances in
understanding,
experimental confirmation of the model has been more difficult to obtain due to
the  difficulty of measuring the surfactant concentration and its dynamics.
Consequently, attention has primarily focused on the evolution of the surface
height profile \cite{Ahmad1972, Bull-1999-SSS, Dussaud-2005-SCI}, with the
location of the leading edge $r_0$ of the surfactant layer
inferred from other dynamics \cite{Bull-1999-SSS}. Within the near-monolayer
regimes for which the model was developed, two experiments have observed
$r_0(t)
\propto t^{0.6 - 0.8}$ spreading behavior on millimetric glycerin films, for
both oleic acid \cite{Gaver-1992-DST} and fluorescently-tagged phosphocholine
(NBD-PC)  \cite{Bull-1999-SSS}.  In experiments  with initial concentrations
$\bar{\Gamma}_0$ of surfactant exceeding 
$\Gamma_c$, spreading behavior consistent with the predicted $r_0(t) \propto
t^{1/4}$ were observed by Dussaud et al.
\inlinecite{Dussaud-2005-SCI} for oleic acid on a
sub-millimetric water-glycerin mixture, and by Fallest et al.
\inlinecite{Fallest-2010-FVS} for NBD-PC on millimeter-thick glycerin. The
latter experiments were able to simultaneously measure both the capillary ridge
and the spatiotemporal dynamics of $\Gamma(r,t)$, to be analyzed in more detail
below.

\begin{figure}
\centerline{\includegraphics[width=\linewidth]{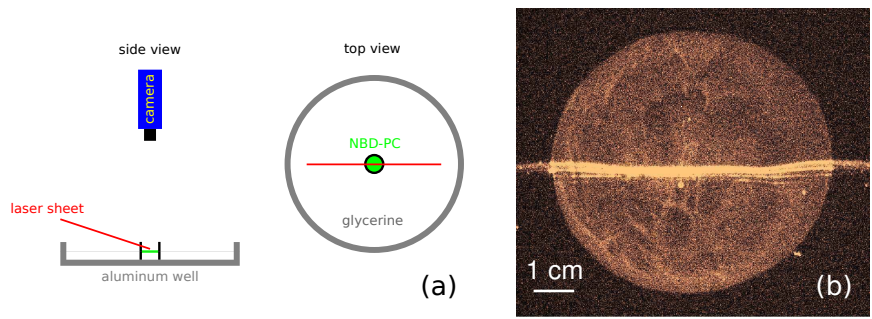}}
\caption{(a) Schematic of apparatus.
(b) Sample image from experiments with initial conditions IC5, with both the
height profile (bright line) and the
fluorescence signal from the NBD-PC lipids (bright disk) at $t^* = 21$~sec.}
\label{f:exp-image}
\end{figure}

In order to verify the model prediction for the time-dependent distribution of
surfactant concentration, we explore stricter tests of the
model than have been performed previously.  Using data from Fallest et al.
\inlinecite{Fallest-2010-FVS} (for which $\Gamma$ is well above $\Gamma_c$), we
make
a detailed comparison
between the model predictions  and the measured
surface height profiles $h(r,t)$  (from a laser line) and measured surfactant
concentration profiles $\Gamma(r,t)$ (from azimuthal averages of fluorescent
intensity at each point $\vec r$). Figure~\ref{f:exp-image} provides a schematic
of the apparatus and a sample image. The well-specified physical parameters
additionally allow us to evaluate the accuracy of the characteristic timescale
predicted for the spreading rate, rather than just the exponent.
While we find approximate agreement in
the spreading exponent and the coincidence of the surfactant leading edge
with the capillary ridge, we also find two significant inconsistencies. First,
there is a mismatch between the characteristic timescale between model and
experiment. Second, the spatial distribution of the surfactant differs
markedly from what is predicted in simulations. To account for the extent to
which these discrepancies might be due to amounts of surfactant well beyond the
monolayer regime, we perform new experiments, modified to allow for the
detection of monolayer concentrations of surfactant ($\Gamma < \Gamma_c$). 
 In these experiments, we observe a distribution of surfactant
which differs from the model predictions. In addition, we find that there
is no spreading capillary ridge, and that the  spreading exponent for the
leading edge of the surfactant is less than $1/10$. This value is well below
predictions of the theory.

The outline of the paper is as follows. In \S\ref{s:experiment}, we provide
details about the experimental setup and the materials used. In
\S\ref{s:model}, we review the Gaver-Grotberg model \cite{Gaver-1990-DLS},
including a discussion of the choice of equation of state relating surfactant
concentration to surface tension, and outline the finite difference method used
for numerical simulation. In \S\ref{s:highgamma}, we describe
comparisons between numerical simulations and experimental results at initial
concentrations well above $\Gamma_c$. As described above, we find partial
agreement but also two inconsistencies when comparing experimental observations
to numerical simulations. In an attempt to address the latter problem, in
\S\ref{s:hybrid} we describe a hybrid model that takes the
experimentally-measured $\Gamma(r,t)$ and uses that quantity (instead of the
model for surfactant evolution) to generate the
evolution of the surface height profile. We find that this provides reasonable
agreement with the experimental data for $h(r,t)$. In particular, the timescale
in the simulations is set by the experimental surfactant distribution, and 
the height profile dynamics therefore evolve according to the
experimentally-observed timescale.
In \S\ref{s:lowgamma}, we report new experiments with
$\Gamma < \Gamma_c$ and find significant disagreement with the model
predictions, as summarized above. We conclude in \S\ref{s:discussion} with a
discussion of the results and their significance.


\section{Surfactant-Spreading Experiments \label{s:experiment}}

In our experiments, we simultaneously record the surface height profile
$h(r,t)$ of the underlying glycerin fluid layer, 
as well as the local fluorescence intensity,
which  corresponds to the   local concentration $\Gamma^*({\vec r}, t)$ of
insoluble lipids (surfactant) spreading on the surface.  The basic apparatus,
described in \inlinecite{Fallest-2010-FVS} and
shown in Figure~\ref{f:exp-image},  consists of an aluminum well, a
black light for exciting the NBD fluorophores, an oblique red laser line to
illuminate the profile of the fluid surface, and a digital camera positioned
directly above the experiment to capture the laser line and fluorescence; this
apparatus is used for the data presented in \S\ref{s:highgamma}.  New
experiments, presented in \S\ref{s:lowgamma},  are optimized to permit
visualization of monolayer concentrations of surfactant. The bottom of the
aluminum well is covered with a plasma-cleaned silicon wafer for
improved reflectivity, and the fluorescent excitation is provided by 467~nm
(blue) LEDs which coincide with the 464~nm absorption peak of the
NBD fluorophore.  A green laser line illuminates the profile of the
fluid surface, so that both it and the fluorescently-emitted light pass through
a band-pass filter centered at the emission peak (531~nm). These improvements
to the optics permit us to collect images of
the spreading dynamics at a framerate of 3~Hz and an integration time of
$\frac{1}{4}$~second, using an Andor Luca R camera optimized for fluorescence
measurements. The signal-to-noise ratio now sets a lower limit of ${\cal
O}(10^{-2}) \Gamma^*_c$ for the detection of surfactant. 

For all experiments, we deposit
1-palmitoyl-2-\{12-[(7-nitro-2-1, 3-benzoxadiazol-4-yl)amino]lauroyl\}
-sn-glycero-3-phosphocholine (abbreviated NBD-PC, from Avanti
Polar Lipids) within a retaining ring which is lifted to begin the spreading
process. This lipid molecule has one 12-carbon chain and one 16-carbon
chain; the NBD fluorophore is attached to the 12-carbon chain.
Experiments are conducted on a layer of  99.5\% anhydrous
glycerin, at an initial depth of $d=0.98 \pm 0.03$~mm, and  held at room
temperature ($22.5 \pm 1.5$ $^\circ$C). The lipids are initially deposited
while dissolved in chloroform, which is allowed to evaporate for at least
$30$~min before the retaining ring is lifted by a motor at $1$~mm/min.
This allows sufficient time for the meniscus to drain before it detaches
from the ring.

\begin{table}
\begin{center}
\begin{tabular}{|c|c|c|}
\hline
Symbol & Interpretation & Value\\
\hline
$\rho$&fluid density, glycerin &$1.26$~g/cm$^3$\\
$\mu$&dynamic viscosity, glycerin&$0.83\pm .03$~Pa$\cdot$s
\cite{Segur-1951-VGA}\\
$D_s$&surface diffusivity,surfactant&$10^{-4}$~cm$^2$/sec
\cite{Sakata-1969-SDM}\\
$\sigma_0$&surface tension, clean glycerin&$63.4\pm 0.3$~dyne/cm
\cite{Wulf-1999-SDS}\\
$\sigma_m$&surfactant-contaminated surface tension&$35.3\pm 0.3$~dyne/cm
\cite{Bull-1999-SSS}\\
$S$&change in surface tension, $S=\sigma_0-\sigma_m$&$28.1\pm.06$~dyne/cm\\
$\Gamma_c$& critical monolayer concentration&$0.3$~$\mu$g/cm$^2$
\cite{Bull-1999-SSS}\\
$H_0$&initial fluid thickness&$0.98\pm .03$~mm\\
$R_0$&lateral dimension&$0.8$~cm or $1.4$~cm (ring radius)\\
\hline
\end{tabular}
\end{center}
\caption{Key dimensional parameters.}
\label{t:ndval}
\end{table}

\begin{table}
\begin{tabular}{|c|c|c|c|c|}
\hline
& $M_L$ ({$\mu$}g) & $R_0$ (cm) & $\bar{\Gamma}_0^*$ $\mu$g/cm$^2$) & \\
\hline
IC1 & 0.85 & $1.4$ & $0.13$ & \multirow{4}{*}{$\bar{\Gamma}_0^* < \Gamma_c$} \\
\cline{1-4}
IC2 & 1.4 & $1.4$ & $0.22$ & \\
\cline{1-4}
IC3 & 1.7 & $1.4$ & $0.26$ & \\
\cline{1-4}
IC4 & 0.6 & $0.8$ & $0.30$ & \\
\hline
IC5 & 18.0 & $1.4$ & $2.73$ & \multirow{2}{*}{$\bar{\Gamma}_0^* > \Gamma_c$} \\
\cline{1-4}
IC6 & 18.0 & $0.8$ & $8.95$ & \\
\hline
\end{tabular}
\caption{The six different initial conditions for the experiments presented in
the paper.}
\label{t:InitCond}
\end{table}

The key material parameters for NBD-PC and glycerin are summarized in
Table~\ref{t:ndval}.
The initial conditions (IC) for the experiment are distinguished by the initial
concentration $\bar{\Gamma}_0^*$ of surfactant deposited within the ring,
defined by
\begin{equation}
\label{gammabar}
\bar{\Gamma}^*_0 \equiv \frac{M_L}{\pi R_0^2},
\end{equation}
 where $M_L$ is the mass of NBD-PC deposited and $R_0$ is the radius of the
ring.  Although a small amount of
surfactant remains on the ring after it has been lifted, we nonetheless use the
nominal
concentration $\bar{\Gamma}_0^*$ to describe the different initial conditions
in Table \ref{t:InitCond}. 
The experiments presented in \S\ref{s:highgamma} all begin from IC6, with 
$\bar{\Gamma}_0^*$ above the critical monolayer concentration $\Gamma_c$.
The experiments of \S\ref{s:lowgamma} employ initial conditions that probe the
monolayer regime (below $\Gamma_c$).

\begin{figure}
\centerline{\includegraphics[width=\linewidth]{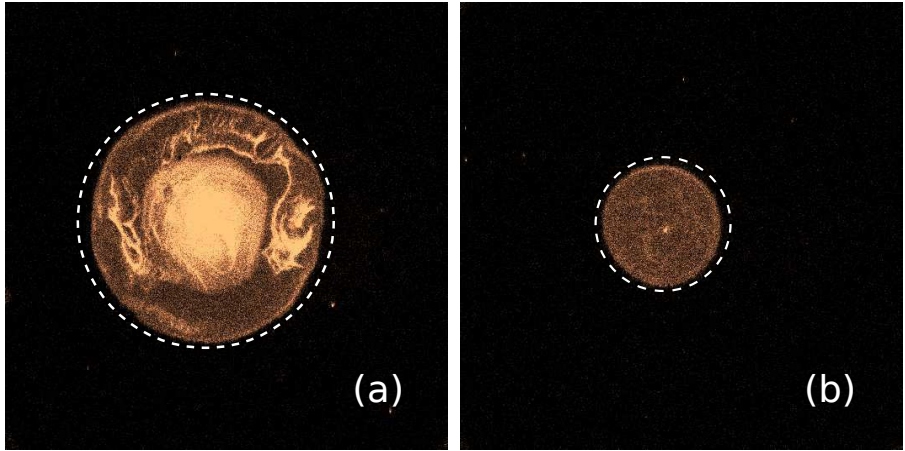}}
\caption{Fluorescence intensity (proxy for surfactant concentration) measured
at $t^* = 5$~sec, for experiments with initial
conditions (a) IC6 and (b) IC4.  The dashed circles (of radii 3.4~cm and 1.1~cm,
respectively), highlight the corrugations in the leading edge.
}
\label{f:IDist}
\end{figure}

Figure~\ref{f:exp-image}b shows a sample image of both the laser
line (measures the height profile $h^*(r)$ and the location $r_M$ of its
maximum), and the fluorescence intensity (measures the surfactant
distribution $\Gamma^*(r)$ after azimuthal averaging and the location $r_0$ of its leading edge).
Figure~\ref{f:IDist} shows sample images of
the surfactant distribution alone,  for a representative
$\bar{\Gamma}^*_0>\Gamma_c$ case and a $\bar{\Gamma}^*_0<\Gamma_c$ case.  In
each image, a sharp 
interface between the surfactant-covered and bare glycerin is readily visible;
the location $r_0$ of this interface is determined by identifying the annulus of
maximum fluorescence intensity gradient. While the surfactant
distribution is uniform for $\bar{\Gamma}_0^* < \Gamma_c$, several
heterogeneities are present when $\bar{\Gamma}_0^* > \Gamma_c$. First, 
the central region contains a greater concentration of surfactant than the
regions closer to the leading edge, an effect that we will explore in more
detail below. Second, there are filamentary patches of high concentration which
also propagate out from the central region, becoming more dilute during the spreading dynamics. We also note that the outer edge of surfactant in both cases has corrugations; in the figure, white circles are imposed to emphasize that the edges are not quite circular. Although the surfactant distributions are never precisely axisymmetric, we nonetheless record the distribution by averaging azimuthally. Moreover, in the model and simulations of the following sections, we assume that the surfactant distributions are axisymmetric.

\section{Model \label{s:model}}

We consider the model derived by  Gaver and Grotberg \inlinecite{Gaver-1990-DLS} for a
single layer of surfactant molecules spreading on a thin liquid film. The model
is a coupled system of partial differential equations for the height $h(r,t)$ of
the fluid free surface and the concentration $\Gamma(r,t)$ (mass per unit area)
of surfactant. We assume  
  axisymmetric spreading,  and the variables are nondimensionalized:
$r=\frac{1}{R_0}r^*$, $t=\frac{1}{T}t^*$, $h=\frac{1}{H_0}h^*$,
$\Gamma=\frac{1}{\Gamma_c}\Gamma^*$, where $^*$ indicates the dimensional
variable.
\small
\begin{subequations}\label{rfulleqn1}
\begin{eqnarray}
h_t+\frac{1}{r}\left(\half
rh^2\sigma(\Gamma)_r\right)_r=\beta\frac{1}{r}\left(\thrd
rh^3h_r\right)_r&-&\kappa\frac{1}{r}\left(\thrd
rh^3\left(h_{rr}+\frac{1}{r}h_r\right)_r\right)_r \label{rfulleqn1h}\\
\Gamma_t+\frac{1}{r}\left(rh\Gamma\sigma(\Gamma)_r\right)_r=\beta\frac{1}{r}
\left(\half rh^2\Gamma h_r\right)_r&-&\kappa\frac{1}{r}\left(\half
rh^2\Gamma\left(h_{rr}+\frac{1}{r}h_r\right)_r\right)_r\nonumber\\&+&\delta\frac{1}{r}
\left(r\Gamma_r\right)_r \label{rfulleqn1g}.
\end{eqnarray}
\end{subequations}
\normalsize
The nondimensional parameter groups $\beta$, $\kappa$, and $\delta$   result
from nondimensionalization, using  values of physical parameters listed in
Table~\ref{t:ndval}. The parameter $\beta=\frac{\rho g H_0^2}{S}\approx 0.42$
balances gravity and Marangoni forces, $\kappa=\frac{\sigma_m
H_0^2}{SR_0^2}\approx 0.019$ is the ratio of the capillary driving forces to the
forces from the surface tension gradient, and
$\delta=\frac{1}{Pe}=\frac{\mu D}{SH_0}\approx 3.0\times 10^{-5}$
represents the surface diffusion of the
surfactant molecules where $Pe$ is the P\'eclet number.  The function
$\sigma(\Gamma)$ expresses the dependence of surface tension $\sigma$ on
surfactant concentration $\Gamma$. It is specified by an equation of state, as
discussed
in the next subsection. The  timescale $T= \frac{\mu R_0^2}{SH_0}\approx
2.0$~sec achieves a balance between the terms on the left side of
(\ref{rfulleqn1}). In \S\ref{modelparams}, we  test this predicted timescale
directly. 

With $\delta>0$, the leading edge of the surfactant distribution is not
precisely defined, since $\Gamma(r,t)>0$ for all $r$ and for all $t>0$.
Nonetheless, although  $\delta$ is very small, it is retained in numerical
simulations,  apart from one case, in which $\delta$ is set to zero in order to
track the leading edge of the surfactant distribution. 

Since the boundary at $r=0$ is not a physical boundary, boundary conditions
there are natural; for large $r$,   the free surface is  undisturbed on the time
scale of the experiment, and the surfactant concentration is expected to be
identically zero: 
\begin{eqnarray}\label{bcsym}
h_r(0,t)=\Gamma_r(0,t)=0,\qquad  h_{rrr}(0,t)=0,\nonumber\\
\lim_{r\to\infty}h(r,t)=1,\qquad\lim_{r\to\infty}\Gamma(r,t)=0.
\end{eqnarray}

A finite difference method is used to simulate (\ref{rfulleqn1}) and is
summarized in the Appendix. The initial condition $h(r,0) $ is chosen to
reflect
the initial height profile in the experiment, as the fluid  meniscus detaches
from the ring, thereby releasing the surfactant to spread across the fluid
surface. The initial distribution $\Gamma_0(r)$ of surfactant within the ring is
unknown, and is varied in the simulations to test its effect on the spreading. 
\begin{eqnarray}\label{numic}
h(r,0)&=&\begin{cases}a\sin\left(2r-\frac{\pi}{3}\right)+(1+a),&\mbox 0\leq
r<\frac{11\pi}{12}\\[6pt]
1,&\frac{11\pi}{12}<r<R_{max},
\end{cases}\nonumber\\
\Gamma(r,0)&=&\begin{cases}\Gamma_0(r),&\mbox 0\leq r<\frac{5\pi}{12}\\[6pt]
0,&\frac{5\pi}{12}<r<R_{max}.
\end{cases}
\end{eqnarray}
In the simulations, we set   $a=0.15$  in order provide a match between the
initial condition and the measured height profiles at later times.
For the simulations, the location $R_{max}$ of the
edge of the domain  is taken large enough that the influence of the
boundary
conditions at $r=R_{max}$ is negligible  over the time of the experiment. We
take
$R_{max}=10$, but display graphs of $h$ and $\Gamma$ over the  smaller domain,
$0\leq
r \leq 7$.

\subsection{Equation of State \label{EoS}}

\begin{figure}
\centerline{\includegraphics[width=0.5\linewidth]{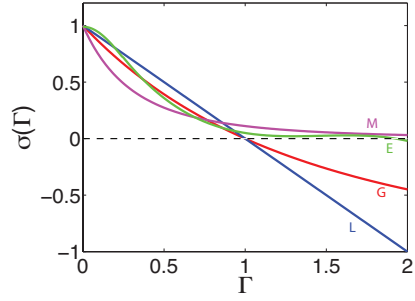}}
\caption{The four equations of state used in the model,  L: linear
(\ref{lineos}), G: Langmuir
(\ref{langeos}), E: measured (\ref{bulleos}), M: multilayer (\ref{meos}).}
\label{f:eos}
\end{figure}

In order to compare the model (\ref{rfulleqn1}) to the results of the
experiments described in \inlinecite{Fallest-2010-FVS}, we need to choose an
appropriate equation of state relating the surfactant concentration $\Gamma$ to
the surface tension $\sigma$.  However, the model is valid
only for a single layer of surfactant molecules ($\Gamma\leq1$) but the
experiments are conducted with initial surfactant concentrations  up to
$\Gamma=30$. In an attempt to extend the model to the regime of the experiment,
we consider different equations of state $\sigma=\sigma(\Gamma)$ that have been
proposed in the literature.  In Figure~\ref{f:eos}, we show the graphs of four
such functions;  we argue below that only the graph labeled {\bf M} is
suitable for modeling the full range of surfactant concentrations we wish to
consider.

The equation of state we seek should have the following properties:
   $\sigma'(\Gamma)<0$,
expressing the effect that increasing  surfactant concentration decreases
surface tension;    $0<\sigma(\Gamma)\leq 1$, since this is the range of
values of surface tension in our  nondimensionalization, with $\sigma(0)=1$. As
can be seen in the figure, only graph {\bf M} has these properties.  

The   linear equation of state
\begin{eqnarray}
\sigma(\Gamma)=1- m\Gamma\label{lineos}
\end{eqnarray}
and has been used widely
\cite{Jensen1993, Matar-1999-SSM, Edmonstone-2004-FSL, Levy-2007-GDT}. This
equation is generally chosen for simplicity; it is also a
reasonable linear approximation to nonlinear equations of state at low
concentration. Note that $\frac{d\sigma(\Gamma)}{d\Gamma}=-m$ is a negative
constant ($m=1$ in the figure). In the case of more than a monolayer of
surfactant,
this equation suggests that the surface tension decreases endlessly which is not
physical, as surfactant concentration beyond a monolayer has little further
effect
in decreasing surface tension.

Bull et al. \inlinecite{Bull-1999-SSS} determined an equation of state for NBD-PC on
glycerin by fitting the data obtained using a tensiometer. Using the
nondimensional parameters in Table \ref{t:ndval}
 ($\Gamma_c\approx 0.3$~$\mu$g/cm$^2$), the corresponding formula for
$\sigma(\Gamma)$ is
\begin{eqnarray}\label{bulleos}
\sigma(\Gamma)=\begin{cases}0.28\cos\left(6.28\Gamma\right)+0.71,& \Gamma<0.25\\
1.26-2.6\Gamma+1.8\Gamma^2-0.41\Gamma^3,& 0.25\leq\Gamma\leq1.67.
\end{cases}
\end{eqnarray}
However, this formula is applicable only for surfactant concentrations below
approximately $2\Gamma_c$
($\Gamma<2$ in Figure~\ref{f:eos}), as $\sigma'(\Gamma) $ decreases
sharply for larger values of $\Gamma$. 

The Langmuir equation of state, used in \inlinecite{Gaver-1990-DLS} and
\inlinecite{Warner:2004}, is
\begin{eqnarray}
\sigma(\Gamma)=\frac{\eta+1}{(1+\Theta(\eta)\Gamma)^3}-\eta\label{langeos}
\end{eqnarray}
where $\Theta(\eta)=\left(\frac{\eta+1}{\eta}\right)^{\frac{1}{3}}-1$ and
$\eta=\frac{\sigma_m}{S}$. When only a small amount of surfactant is introduced,
a large change in the surface tension occurs and as the surfactant comes close
to saturation (a monolayer) then adding more surfactant does not alter the
surface tension very much. However, the range of $\sigma(\Gamma)$ is $[-\eta,
1]$ rather than $[0,1]$. The multiple layer equation of state used by Borgas and Grotberg
\inlinecite{Borgas-1988-MFT} is related to the Langmuir equation of state:
\begin{eqnarray}\label{meos}
\sigma(\Gamma)=(1+\eta\Gamma)^{-3}
\end{eqnarray}
This formulation is based on  properties of
surface tension discussed by Sheludko \inlinecite{Sheludko:1966} and by an experimental
fit by Foda and Cox
\inlinecite{Foda-1980} who worked with an oil layer on water. In addition,
$\sigma(\Gamma)$ remains positive at large $\Gamma$, allowing us to
simulate much higher concentrations of surfactant. Therefore, we
use Eq.~(\ref{meos}) for the simulations.

\section{High Surfactant Concentration ($\bar{\Gamma}_0^* > \Gamma_c$)
\label{s:highgamma}}

The $r_0(t) \propto t^{1/4}$ spreading behavior predicted by (\ref{rfulleqn1})
has already been observed in prior experiments with oleic acid on a
water-glycerin mixture \cite{Dussaud-2005-SCI} and NBD-PC on glycerin
\cite{Fallest-2010-FVS}. In this section, we make a more detailed comparison
between the results from  experiments and numerical simulations, using
data from Fallest et al. \inlinecite{Fallest-2010-FVS}. While we are able to obtain
reasonable agreement in the height profile shapes, a comparison of the
dynamics (\S\ref{modelparams}) requires that we adjust the timescale. In
\S\ref{compprofile}, we show a significant
discrepancy between the observed distribution of surfactant and the prediction
from simulations, even though the location and time evolution of the leading
edge of the surfactant layer agree well, as detailed in \S\ref{spreadingexp}. We
also describe attempts to capture the experimentally observed surfactant
distribution by varying the initial distribution $\Gamma_0(r)$ in the
simulations.

\subsection{Timescale \label{modelparams}}

\begin{figure}
\centerline{\includegraphics[width=0.5\linewidth]{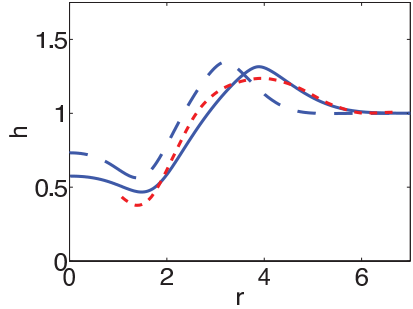}}
\caption{Surface height profiles $h(r)$ measured at  $t^* = 5$~sec in
the experiment (short dashed lines, red), compared to numerical solutions at
nondimensional times. Long dashed lines: $t = 2.5$ (uses calculated $T=2.0$
sec); Solid line: $t = 5$ (uses better-fitting $T = 1.0$~sec). Data are from
Fallest et al. \inlinecite{Fallest-2010-FVS}, with initial condition IC6.}
\label{f:thalf}
\end{figure}

In  order to compare the model and experiment, we convert the simulation results
from dimensionless time $t$ to dimensional time $t^*$ using the
relation $t=\frac{t^*}{T}$,
where $T \equiv \frac{\mu R_0^2}{SH_0}\approx 2.0$~sec is calculated
from the dimensional parameters  listed  in Table~\ref{t:ndval}.
Previous simulations \cite{Gaver-1992-DST, Dussaud-2005-SCI}  have treated the
lengthscale $R_0$ as a free parameter, effectively adjusting the timescale to
agree with the experimental observations. However, for the experiments
analyzed here, the ring radius $R_0=0.8$~cm is known and consequently the time
scale $T$ is determined, with no free parameters.

In Figure~\ref{f:thalf},  we observe that the simulated height profile and the
experimental data  are inconsistent at $t^*=5$~sec if the determined value
$T=2.0$~sec is used: neither the peak location nor its width are in agreement
with the model. However, if we use
$T=1.0$~sec instead of $T=2.0$~sec, thereby comparing the simulation at the
later time $t=5$ to the same experimental data at $t^*=5$~sec, then both the
position and width 
of the ridge are in approximate agreement between the model and experiment. This
agreement between simulation and experiment using the timescale $T=1.0$~sec is
observed to hold for all times beyond an initial transient.  Although an 
explanation for the discrepancy between times scales awaits further
investigation, we use the empirically-determined $T=1.0$~sec as the timescale
for the remaining 
comparisons in the paper.

\subsection{Height Profile and Surfactant Distribution  \label{compprofile}}

\begin{figure}
\centerline{\includegraphics[width=\linewidth]{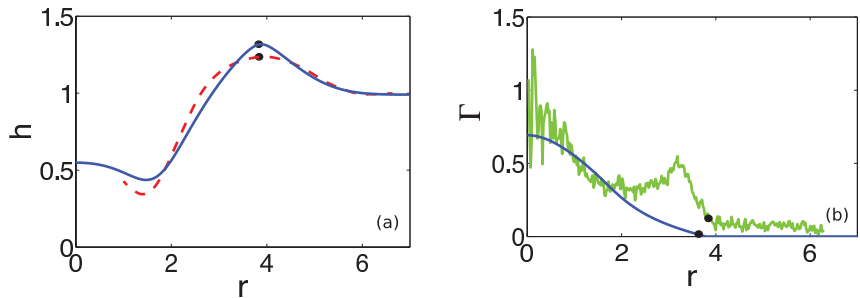}}
\caption{
(a) Surface height profiles $h(r)$ measured at
$t^*=5$ seconds in the experiment (dashed lines) and $t=5$   in
numerical solution (solid line); the location $r= r_M$ of the peak   is
marked with a $\bullet$.
 (b) The corresponding surfactant concentration profiles $\Gamma(r)$ in the
experiment (noisy) and numerical solution (smooth); the location $r=r_0$ of the
leading edge of the surfactant is marked with a $\bullet$. Data are from
Fallest et al. 2010, with initial condition IC6.}
\label{f:profile1}
\end{figure}

In Figure~\ref{f:profile1} we compare the simulated and measured surface
profiles and
surfactant distributions, using the model parameters described in
\S\ref{modelparams} and experimental data from Fallest et al. \inlinecite{Fallest-2010-FVS}. As
can be seen in Figure~\ref{f:profile1}, the height profiles $h(r,5)$ are in
approximate
agreement: locations of  the  maximum
and minimum are in approximate agreement between simulation and experiment, and
the overall shapes are
similar. In contrast, the measured surfactant distribution has quite a different
shape from the distribution  predicted by the simulations. While the model
predicts a smooth decrease in 
$\Gamma(r,5)$ away from the central peak, experiments instead show an extended
plateau over which the surfactant concentration is nearly constant, and
 which appears to be drawn out of a reservoir, near the peak concentration at
$r=0$. For longer times, the plateau extends and decreases in height. These
features do not appear in the numerical simulations.

\begin{figure}
\centerline{\includegraphics[width=\linewidth]{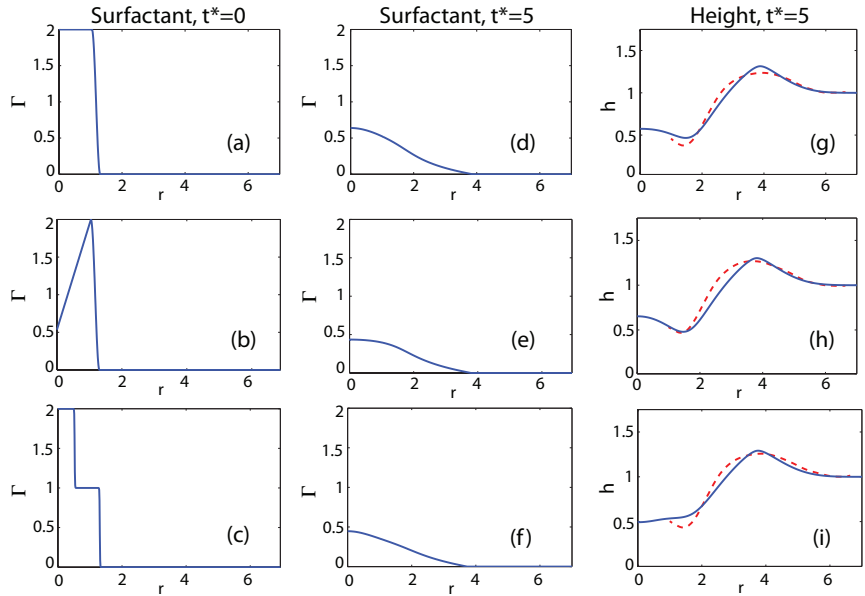}}
\caption{(a-c) Three different surfactant initial conditions, with the
resulting (d-f) surfactant concentration profiles $\Gamma(r)$ and (g-i) surface
height profiles $h(r)$ from numerical solutions at $t=5$. The experimental
height profiles at $t^*=5$\ seconds are the same in each case, and are shown as
dashed lines.}
\label{f:surfic}
\end{figure}

In the experiment,  the total surfactant mass is known, but its    initial
spatial distribution within the retaining ring is not measured. 
Consequently, there is some uncertainty about the appropriate initial condition
$\Gamma(r,0)=\Gamma_0(r)$.
We explore whether the choice of $\Gamma_0(r)$ could change the
simulations enough to replicate the  expanding
plateau in the experimentally observed  surfactant distributions.  We tested
three  different functions $\Gamma_0$, shown in
Figure~\ref{f:surfic}: (a) Uniform distribution;
(b) Surfactant more concentrated near the retaining ring;
(c) Step distribution. The initial free surface height profile $h(r,0)$ is the
same in each case, given by (\ref{numic}). 

The results of these simulations at time $t^*=5$~sec  (after short-time
transients 
have died out) are shown in
Figure~\ref{f:surfic}. In the middle column, we observe that the distribution
$\Gamma$ of surfactant does not exhibit a plateau for any of the initial
conditions, while in the final column we see that the height profiles show broad
agreement  with the experiment in each case.

\subsection{Spreading Exponent \label{spreadingexp}}

\begin{figure}
\centerline{\includegraphics[width=\linewidth]{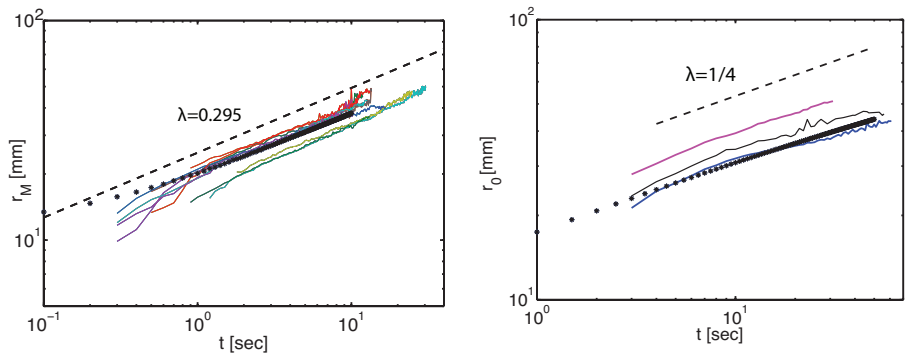}}
\caption{Comparison between spreading rates of (a) the location of the maximum
of the surface height profile $h(r)$ and (b) the location of the leading edge of
surfactant in both experiment and simulation. Colored lines are from
experiments with IC6 adapted from Fallest et al. 2010, black dots are
from simulations. Dashed lines are (a) comparison to best-fit $t^\alpha$ with
$\alpha = 0.295$ and (b) comparison to $t^{1/4}$.}
\label{f:spread}
\end{figure}

In spite of the disagreements above, we observe that the spreading dynamics of
the model and experiment are in good agreement when the artificial choice of
$T=1$~sec  is used. The surface diffusion term $\delta\Gamma_{xx}$ in
(\ref{rfulleqn1h}) smooths the surfactant profile, and  guarantees
$\Gamma(r,t)>0$ for  all   $r\geq 0$ and $t>0$. This means the surfactant
distribution has no clearly defined leading edge. Since $\delta\approx
10^{-5}$ has a very small effect,  we take $\delta=0$ in these simulations. 
 The surfactant
distribution is then supported at each $t>0$ on a bounded interval
$0\leq r\leq r_0(t)$,   and
the leading edge $r=r_0(t)$ can be tracked using the numerical scheme described
in \inlinecite{Peterson-2012-SSS} with $\beta\kappa>0$.
In Figure~\ref{f:spread}(b), $r_0(t)$ is shown as a dotted line; on the log-log
plot
the numerical solution is  compared to the experimental results and to the
analytic form $r_0(t)\sim t^{1/4}$ derived from the similarity solution  of
\inlinecite{Jensen-1992-ISS}, in which $\beta=\kappa=\delta=0$.

We also tracked the capillary ridge $r=r_M(t)$,  where the height profile
$h(r,t)$ has a maximum. In Figure~\ref{f:spread}(a) we show the numerical
solution
as a dotted line, and compare to experimental results, with an approximate slope
shown with a dashed line.  The data for the leading edge of the surfactant
agrees
with the $t^{\frac{1}{4}}$ prediction of the model.  The  capillary
ridge  moves faster as it catches up to the surfactant leading edge; it  is best
fit by $t^{0.295}$ over the duration of the experiment.

Note that for $\beta=\kappa=\delta=0$ in the model,  the fluid surface
experiences a discontinuity at  the leading edge $r=r_0(t)$ of the surfactant,
and the fluid
is undisturbed ahead of this front.  It is worth noting that with  $\delta=0$,
the surfactant distribution still has a finite extent when $\beta$ or  $\kappa$
is non-zero, but  the disturbance of the film does not. In fact, the surface
tension gradient induced at the leading edge $r=r_0(t)$ of the surfactant
generates fluid motion ahead of $r_0$.
 

\section{Hybrid Model \label{s:hybrid}}

Due to the failure of the model equations (\ref{rfulleqn1}) to capture
the spatial distribution of the surfactant, as discussed in \S\ref{s:highgamma},
we
consider whether there is a fundamental modeling problem with the equation
(\ref{rfulleqn1g}) for the evolution of $\Gamma(r,t)$. Given that the height
profile $h(r,t)$ evolves in a quantitatively reasonable way, we
leave (\ref{rfulleqn1h}) intact, but take
$\Gamma(r,t)$  from the experiment and use it to determine $h(r,t)$ numerically.

The experimental data for $\Gamma$ are noisy, and occur at discrete
times; we must smooth and interpolate the data for use in the numerical
scheme to determine $h$.
The first step is to smooth the experimental data at each recording time. This
is achieved 
using the  MATLAB \textit{smooth} function, which performs a moving spatial
averages over a specified span of data points; we found a low-pass filter with a
span of 21 points to be effective.

The experimental  data are recorded at one second intervals; however, the
numerical method requires the time step $\Delta t$ to be on the order of
${\cal O}(10^{-3})$  for stability. To remedy this inconsistency, we interpolate
the
smoothed surfactant profiles to obtain functions that can be used to represent
the  surfactant concentration $\Gamma(r,t)$ at all values of $r$ and $t$.

\begin{figure}
\centerline{\includegraphics[width=\linewidth]{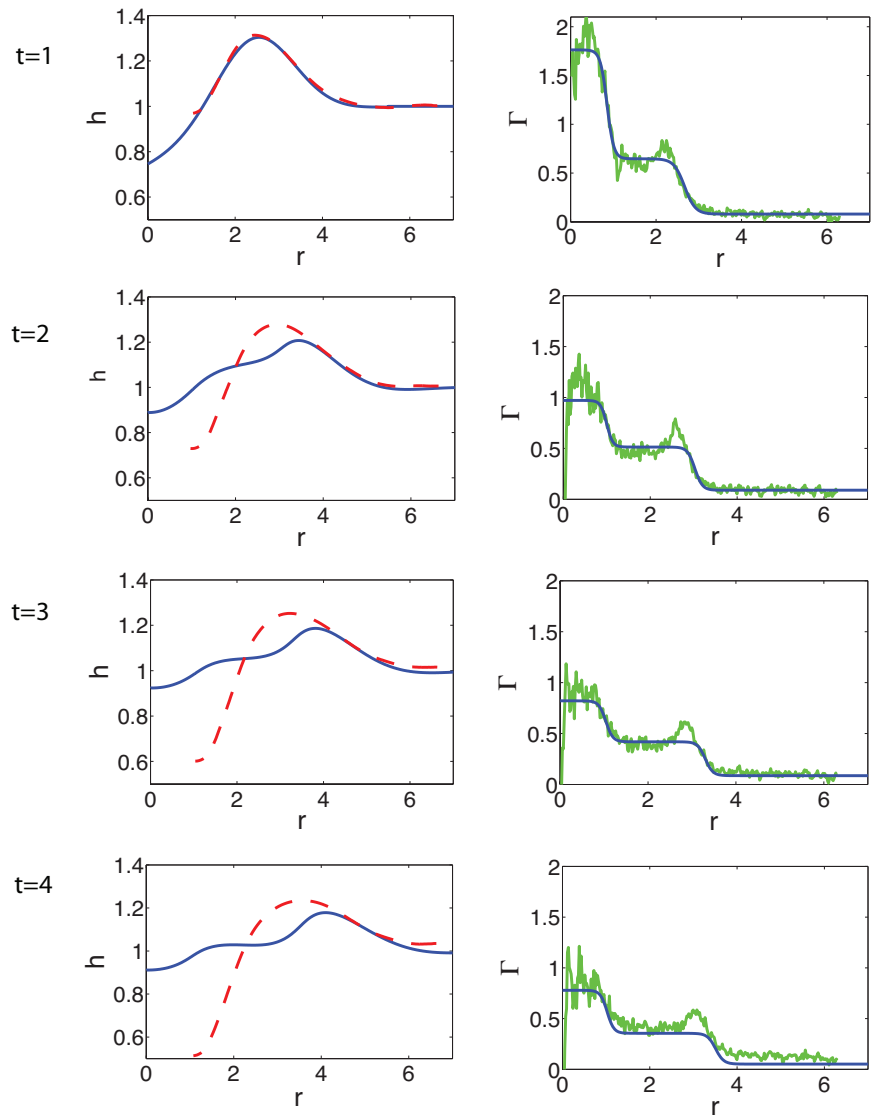}}
\caption{(a-d) Comparison of the surface height profiles $h(r)$ from the
numerical
simulations (solid blue) and experimental data (dashed red). (e-h) The
corresponding
surfactant profiles from the experiment, smoothed data (smooth blue) and raw
data (noisy green). The profiles are shown at $t^*=1,2,3,4$ sec. }
\label{f:exp4}
\end{figure}

First, we use the nonlinear fit function, \textit{nlinfit} in MATLAB to fit the
smoothed surfactant data at each experimental recording time to a function with
a graph consistent with the two step structure of the experimental surfactant
distributions. As observed in Figure~\ref{f:exp4}, the experimental surfactant
distribution   $\Gamma$ is roughly constant in each of the two steps. With this
in mind, we use the function
\begin{eqnarray}
\Gamma(r,j)=a_1^j-a_2^j\tanh\left(\frac{r-a_3^j}{a_4^j}
\right)-a_5^j\tanh\left(\frac{r-a_6^j}{a_7^j}\right)
\end{eqnarray}
to fit the surfactant profile at each time $t^*$ for which we have data, where
$t^*=j, j=1,2,\dots, 10$~sec. This procedure generates coefficients
$a_1^j,\dots, a_7^j$.  

Next we create polynomial functions, $a_i^*(t^*), i=1\dots 7$ from the discrete
values $a_i^j,  j=1,\dots 10$ using
\textit{polyfit} in MATLAB, but with a restriction that the mass of the
surfactant must be conserved. Since the numerical code requires nondimensional
time 
$t=0.5t^*$, we define  $a_i(t)=a_i^*(2t), i=1\dots 7$.
These functions define the surfactant concentration profiles that approximate
the experimental data:
 \begin{eqnarray}\label{gfit}
\Gamma(r,t)=a_1(t)-a_2(t)\tanh\left(\frac{r-a_3(t)}{a_4(t)}
\right)-a_5(t)\tanh\left(\frac{r-a_6(t)}{a_7(t)}\right).
\end{eqnarray}

We update the height profile using the finite difference scheme used in
\S\ref{s:highgamma} with boundary conditions (\ref{bcsym}) and the surfactant
concentration profile using (\ref{gfit}): 
\begin{subequations}
\begin{eqnarray}
h_j^{n+1}=h_j^n&+&\Delta t\frac{1}{r_j\Delta
r}\left(\mathcal{F}_{j+\half}^{n+1}-\mathcal{F}_{j-\half}^{n+1}\right)\\
\Gamma_j^{n+1}=a_1(t^{n+1})&-&a_2(t^{n+1})\tanh\left(\frac{r-a_3(t^{n+1})}{
a_4(t^{n+1})}\right)\nonumber\\&-&a_5(t^{n+1})\tanh\left(\frac{r-a_6(t^{n+1})}{a_7(t^{n+1})}
\right).
\end{eqnarray}
\end{subequations}
The flux functions $F_j^n$ are described in the appendix.

In  Figure~\ref{f:exp4}, we show simulation results using initial condition
(\ref{numic})
and parameters $\beta=0.5,\; \kappa=0.5$. Note  that this value of $\kappa$ is
larger than the value suggested by the nondimensional grouping.  This   is in
order 
  to smooth the height profile, which otherwise would develop multiple
persistent ridges, due to the steep gradient in the surfactant concentration.  
Multiple ridges are observed at very early time in the simulations discussed in
\S~\ref{compprofile} but then the gradient in the surfactant is quickly
smoothed. In this case, the ridge in the height profile from the surfactant
gradient and the one from the initial condition (due to the lifting of the ring)
combine at very early time and then propagate as one.  
By contrast, in the experiment the surfactant distribution does not experience
this smoothing and consequently when the experimental values for the  surfactant
concentration  are input into the height  equation a second ridge develops and
persists.  By increasing the capillarity parameter $\kappa$, the height profile
is smoothed, and the numerical height profile becomes more similar to
the experimental height profile. 

The surface height profile simulated using this hybrid model exhibits
an evolution and shape that is comparable to the experiment.   We conclude that
the equation modeling surfactant molecule motion through passive transport by
the surface fluid is fundamentally flawed, and is missing some physical or
chemical properties that would generate the surfactant distributions observed in
the experiment. 


\section{Monolayer Surfactant Concentration ($\bar{\Gamma}_0^* < \Gamma_c$)
\label{s:lowgamma}}

Because the original model (\ref{rfulleqn1}) was developed for use with
monolayer concentrations of surfactants ($\bar{\Gamma}_0^* < \Gamma_c$), we 
conduct new experiments in this regime. In addition, these experiments
help elucidate the discrepancies between model and experiment at the higher
concentrations. Performing experiments at lower surfactant
concentrations requires
improvements of our earlier techniques (see \S\ref{s:experiment}) in order to
visualize lower surfactant concentrations.  We perform experiments
starting from four different initial concentrations, IC1-IC4 in Table
\ref{t:InitCond}, all of which result in similar spreading dynamics, described
below. In no case do we find that the agreement with the model is improved over
the $\bar{\Gamma}_0^* > \Gamma_c$ case: height profiles, surfactant
distribution, and the spreading dynamics all significantly disagree with the
model predictions.

\subsection{Height Profile and Surfactant Distribution \label{compprofilelow}}

\begin{figure}
\centerline{\includegraphics[width=0.7\linewidth]{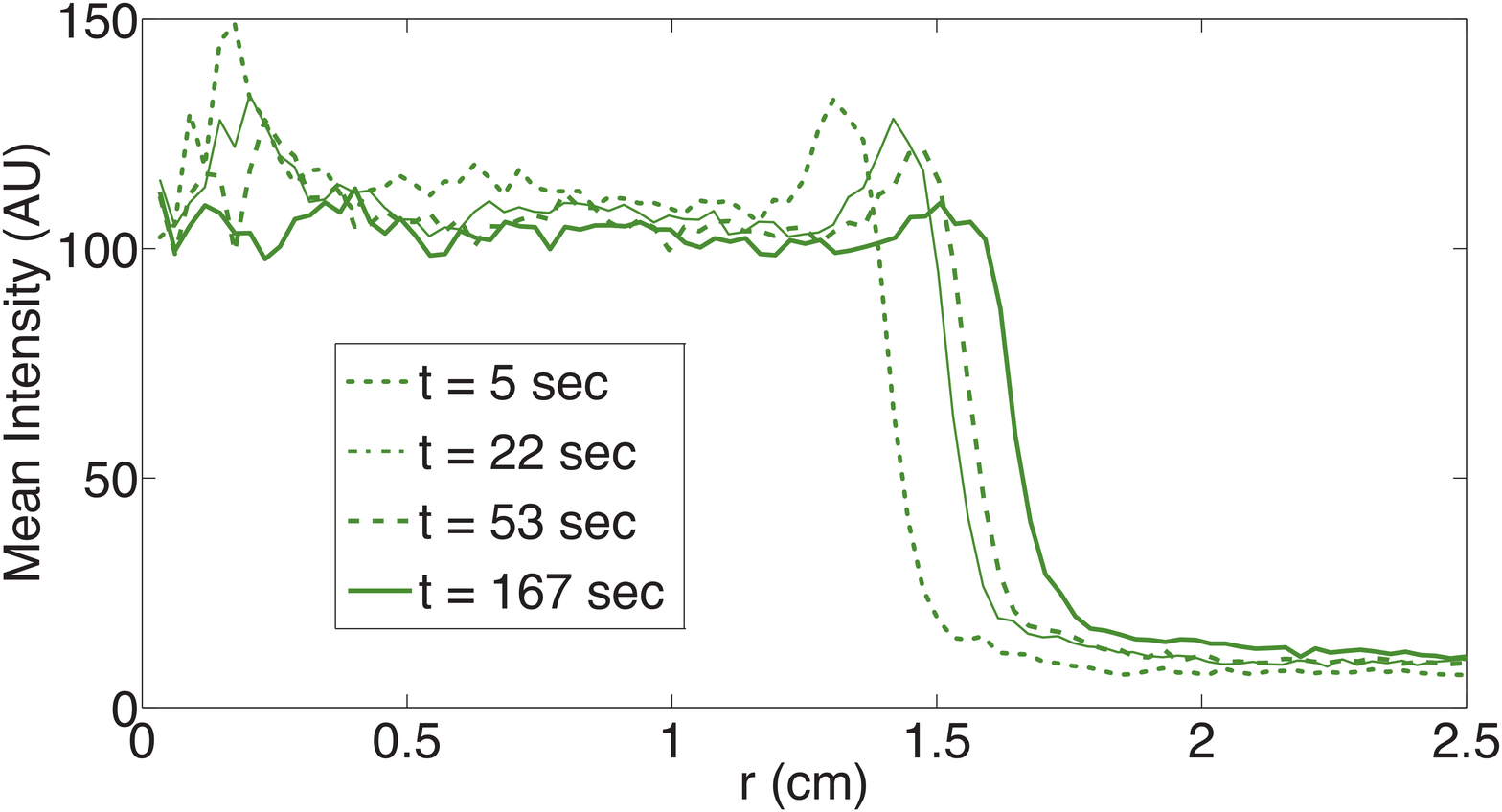}}
\caption{Azimuthally-averaged fluorescence intensity at a distance $r$ from
the center of the surfactant region, shown at representative times, taken from
an experiment  with IC4 ($\bar{\Gamma}_0^* < \Gamma_c$).
}
\label{f:fluoro}
\end{figure}

As illustrated in Figure~\ref{f:IDist}, and shown quantitatively in
Figure~\ref{f:fluoro}, the spreading region has an approximately uniform
surfactant distribution throughout the lipid-covered area. The
leading edge, located at $r_0$, exhibits a sharp interface (approximately
$0.5$~mm wide, neglecting azimuthal corrugations)
that does not broaden as the surfactant spreads outward. Instead, the overall
concentration decreases throughout the lipid-covered region. These 
observations are in disagreement with the model, which predicts a
monotonically decreasing profile (see Figure~\ref{f:profile1}) with a gradual
and broadening transition to $\Gamma=0$ at $r_0$.

\begin{figure}
\includegraphics[width=\linewidth]{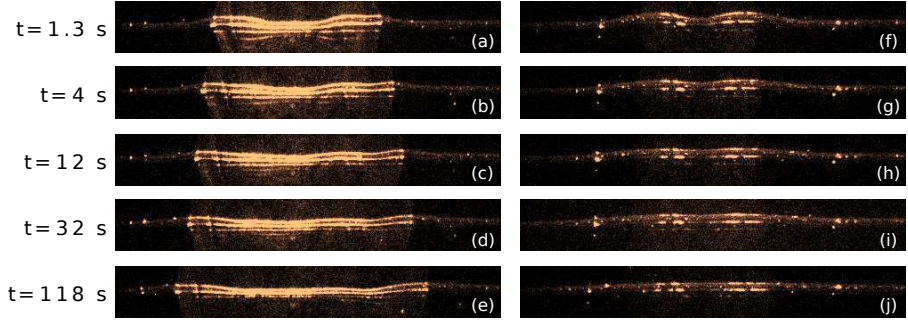}
\caption{Images comparing the laser profile for two experiments with IC5
(a-e) ($\bar{\Gamma}_0^* > \Gamma_c$) and IC2 (f-j) ($\bar{\Gamma}_0^* <
\Gamma_c$)
at five different
times. The uppermost bright line in each image is the reflection from the air glycerin
interface, and is used to measure the free surface height profile $h(t)$.
}
\label{f:MRidge}
\end{figure}

Figure~\ref{f:MRidge} provides a comparison
between the behavior of the fluid ridge in two regimes. For $\bar{\Gamma}_0^*
> \Gamma_c$, the
glycerin is pushed out by the surface tension gradient, and forms a capillary
ridge at the surfactant edge (panels f-j). As already discussed in
\S\ref{compprofile}, these two features travel outward together.
This is consistent with simulations at high or low surfactant concentrations,
that show a fluid ridge propagating outwards due to  Marangoni forces.
In contrast, the experimental data in the $\bar{\Gamma}_0^*< \Gamma_c$ regime
(panels a-e) show an initial ridge (due to the meniscus at the ring) collapsing
rapidly in place, and no discernible ridge propagating outwards.

\subsection{Spreading Exponent \label{spreadingexplow}}

\begin{figure}
\centerline{\includegraphics[width=\linewidth]{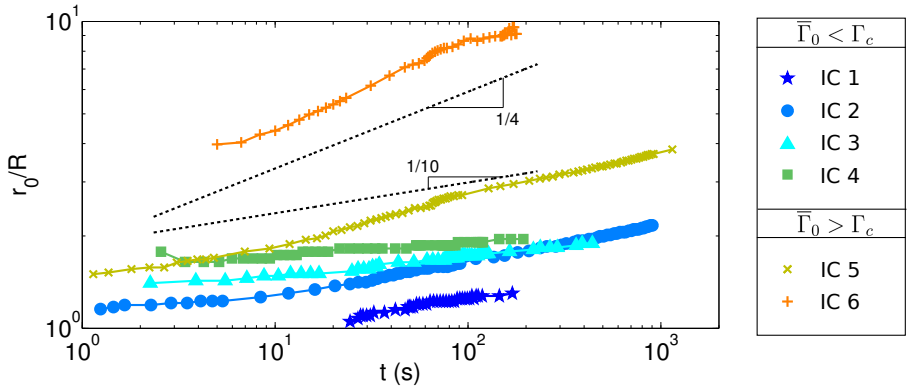}}
\caption{ The spreading dynamics of the location of the leading edge of the
fluorescence intensity ($r_0$), scaled by the ring radius $R_0$. The dashed
 lines corresponding to $t^{1/4}$ and $t^{1/10}$ spreading behavior are shown
for comparison.  }
\label{f:RvsT}
\end{figure}

 Given the significant differences between the observed and modeled $h(r,t)$ and
$\Gamma(r,t)$, it is unsurprising that we
observe the spreading dynamics to be quite different as well. In
Figure~\ref{f:RvsT}, we plot the position of the leading edge $r_0$ as a
function of time,  for experiments in both the monolayer regime of IC1-IC4 and
for
higher initial concentrations (IC5-IC6). For $\bar{\Gamma}_0^* < \Gamma_c$
(IC1-IC4) the dynamics all follow  a form $r_0 \propto t^{\alpha}$, with $\alpha
\lesssim 1/10$. Remarkably, this is reminiscent of Tanner's law for fluid
spreading on a solid \cite{Tanner1979}. For slightly larger values of
$\bar{\Gamma}_0^*$ (IC5), we observe faster surfactant spreading dynamics
initially, but  $r_0 \propto t^{\alpha}$ with 
$\alpha \approx 1/10$ at later times. This
slow-spreading regime was not reached in the runs with much larger values of
$\bar{\Gamma}_0^*$ (IC6), for which $\alpha$
remains close to   $\alpha=1/4$, as predicted in   \cite{Jensen-1992-ISS}. The
decrease in
$\alpha$ for low surfactant concentrations suggests a transition in the dynamics
which is not covered by the model equations. The differing prefactors
for the various runs is probably a result of variations in $\bar{\Gamma}_0^*$
due to
some lipid molecules remaining on the ring after it lifts off, an effect that is
more significant at lower concentrations.

\section{Discussion \label{s:discussion}}

Current mathematical models that describe the dynamics of
the free surface of a thin fluid layer subject to forces induced by variations
in surface tension contain two key assumptions: (1) lubrication theory is
valid, and (2) surfactant molecules are primarily passively-transported
by the fluid motion along the surface, along with negligible molecular
diffusion. However, the dependence of surface
forces on local variations in surfactant concentration is not completely
settled, especially for larger concentrations. Due to the coupling between the
motion of the underlying fluid and the spreading of surfactant molecules, it is
crucial to compare results from simulations and experiments for both fluid
motion (through surface deformations) and the dynamics of surfactant
distribution.

In this paper, we compare model predictions to experiments that include the
simultaneous visualization of the fluid height profile and the distribution
of surfactant,  both above and below the critical monolayer concentration
$\Gamma_c$. In both cases, we find serious inconsistencies between the model
and the experiments. The aspect ratio for the experiments is
$\frac{H_0}{R_0} \approx 0.1$, which justifies the use of the lubrication
approximation; we have not verified the magnitude of the vertical velocity
profile. 
At all initial concentrations, both above and below
$\Gamma_c$, the spatial distribution of surfactants does not follow the smooth,
monotonically-decreasing profiles predicted by the model. At low surfactant
concentrations ($\bar{\Gamma}_0^* < \Gamma_c$, for which the models were
originally developed), the distribution is highly-uniform with a sharp
interface at the leading edge. Second, spreading occurs much more slowly than is
predicted by the model. For all experiments with $\bar{\Gamma}_0^* < \Gamma_c$,
the spreading dynamics of the leading edge approximately follow a power law
  $r_0(t) \propto t^\alpha$,  with $\alpha \lesssim 1/10$. This is
significantly smaller than the $\alpha = 1/4$ predicted by the natural
scaling in the model. Interestingly, this exponent is also
markedly different from the exponent of $1/2$ to $3/4$ observed by
Gaver and Grotberg \inlinecite{Gaver-1992-DST} for oleic acid and by Bull et al. \cite{Bull-1999-SSS} for
NBD-PC.
In the former case, the measurement technique relies on the
model for interpretation of experimental results, while in the later the $2$~mm
fluid layer thickness may be large enough that deviations from the lubrication
approximation are significant. For $\bar{\Gamma}_0^* >\Gamma_c$, even
though $\alpha \approx 1/4$,  there is a mismatch by a factor of two between
the timescales predicted in the model and observed in experiment. These
inconsistencies are not resolved by changing assumptions concerning the initial
distribution of surfactant. Moreover, if a measured $\Gamma(r,t)$ is
incorporated directly into the lubrication model, then the timescale issue is no
longer present, as the spreading ridge is simply driven by the spreading
surfactant.

 One untested assumption is the functional form
of the equations of state which have been considered to date. The lack of
spreading ($\alpha \lesssim 1/10$ for very low
concentrations) might indicate  that the assumed $\sigma(\Gamma)$ equation of
state is inadequate.  If there
were a value of $\Gamma$ below which there were no longer a significant surface
tension gradient, a lack of spreading would be expected. In fact, in static
surface-pressure measurements of diolein, oleyl alcohol, and lecithin on water,
such an effect has been observed \cite{Henderson1998}. Future work to make
similar measurements for NBC-PC, potentially locating a second transition point,
may clarify the reason for the reduction in spreading. Another possibility is
that the passive transport model for surfactant distribution on the free surface
is missing one or more effects that influence on the dynamics of insoluble
surfactant spreading on thin liquid films.

\section{Acknowledgments}

We are grateful for support from the National Science Foundation under grant
number  DMS-0968258 and Research Corporation under grant number 19788. In addition, we wish to thank Rachel
Levy for valuable conversations concerning surfactant spreading.


\clearpage
\section*{Appendix}

 We summarize the finite difference method used to generate numerical results
for system (\ref{rfulleqn1}). Simulations are conducted on a large interval
$0\leq r\leq R$.  For simplicity, we consider uniformly distributed grid points
$r_j=j\Delta r, j=1,\dots,N$,  where $\Delta r=R/N$.
At each time $t_n=n\Delta t, \ n\geq 0$, let $h_j^n\approx h(r_j, n\Delta t)$
and $\Gamma_j^n\approx\Gamma(r_j,n\Delta t)$.
We use the standard notation for spatial averages of $u_j^n=u(r_j,t_n)$,
\begin{eqnarray}
\bar{u}^n_{j+\half}\equiv\frac{u_{j+1}^n+u_j^n}{2}.
\end{eqnarray}
The numerical method is an implicit finite difference scheme in conservative
form:
\begin{subequations}\label{rnumerical}
\begin{eqnarray}
h_j^{n+1}&=&h_j^n+\Delta t\frac{1}{r_j\Delta
r}\left(\mathcal{F}^{n+1}_{j+\half}-
\mathcal{F}^{n+1}_{j-\half}\right)\label{rnumericalh}\\
\Gamma_j^{n+1}&=&\Gamma_j^n+\Delta t\frac{1}{r_j\Delta
r}\left(\mathcal{G}^{n+1}_{j+\half}- \mathcal{G}^{n+1}_{j-\half}\right).
\label{rnumericalg}
\end{eqnarray}
\end{subequations}
The fluxes inherit the structure of the PDE system (\ref{rfulleqn1}) (note that
we have dropped the $n+1$ superscript)
\begin{subequations}
\begin{eqnarray}
\mathcal{F}_{j+\half}&=&\half\mathcal{F}_{j+\half}^1+\thrd\beta\mathcal{F}_{
j+\half}^2+\thrd\kappa\mathcal{F}_{j+\half}^3\\
\mathcal{G}_{j+\half}&=&\mathcal{G}_{j+\half}^1+\half\beta\mathcal{G}_{j+\half}
^2+\half\kappa\mathcal{G}_{j+\half}^3+\delta\mathcal{G}_{j+\half}^4.
\end{eqnarray}
\end{subequations}
The individual  fluxes  are expressed as
\begin{subequations}\label{flux}
\begin{eqnarray}
\mathcal{F}_{j+\half}^1&=&\bar{r}_{j+\half}\left(\bar{h}_{j+\half}
\right)^2\sigma'\left(\bar{\Gamma}_{j+\half}\right)\frac{\Gamma_{j+1}
-\Gamma_j}{\Delta r}\\
\mathcal{F}_{j+\half}^2&=&\bar{r}_{j+\half}\left(\bar{h}_{j+\half}
\right)^3\frac{h_{j+1}-h_j}{\Delta r}\\
\mathcal{F}_{j+\half}^3&=&\bar{r}_{j+\half}\left(\bar{h}_{j+\half}
\right)^3\frac{\mathcal{E}_{j+2}-\mathcal{E}_{j}}{2\Delta r}\\
\mathcal{E}_{j+2}&=&\frac{h_{j+3}-2h_{j+2}+h_{j+1}}{\Delta
r^2}+\frac{1}{\bar{r}_{j+\frac{3}{2}}}\frac{h_{j+2}-h_{j+1}}{\Delta r}\\
\mathcal{G}_{j+\half}^1&=&\bar{r}_{j+\half}\bar{h}_{j+\half}\bar{
\Gamma}_{j+\half}\sigma'\left(\bar{\Gamma}_{j+\half}\right)\frac{\Gamma_{
j+1}-\Gamma_j}{\Delta r}\\
\mathcal{G}_{j+\half}^2&=&\bar{r}_{j+\half}\left(\bar{h}_{j+\half}
\right)^2\bar{\Gamma}_{j+\half}\frac{h_{j+1}-h_j}{\Delta r}\\
\mathcal{G}_{j+\half}^3&=&\bar{r}_{j+\half}\left(\bar{h}_{j+\half}
\right)^2\bar{\Gamma}_{j+\half}\frac{\mathcal{E}_{j+2}-\mathcal{E}_{j}}{
2\Delta r}\\
\mathcal{G}_{j+\half}^4&=&\bar{r}_{j+\half}\frac{\Gamma_{j+1}-\Gamma_j}{
\Delta r}.
\end{eqnarray}
\end{subequations}
In the hybrid model of \S\ref{s:hybrid}, the values of $\Gamma_j^n$ are derived
from the experimental data, as described in that section.  The height profile
evolution can then be computed from equation \eqref{rnumericalh} and the
companion equation \eqref{rnumericalg} is discarded.

\end{document}